\documentclass[a4paper]{appolb}

\usepackage{graphicx}
\usepackage{pgfplotstable}
\usepackage{amsmath}
\usepackage[mode=buildnew]{standalone}
\usepackage{listings}
\usepackage{relsize}
\usepackage[caption=false,
  justification=centerfirst,
  singlelinecheck=false]{subfig}

\providecommand{\e}[1]{\ensuremath{\times 10^{#1}}}

\newlength{\singlecolumnfigurewidth}
\newcommand{\dl}{\Delta l}

\newcommand{\tev}{\widetilde{\textbf{e}}}
\renewcommand{\e}{\mathbf{e}}
\newcommand{\R}{\bar{R}}
\newcommand{\te}{\widetilde{e}}
\newcommand{\tzu}{\widetilde{z_u}}
\newcommand{\tzd}{\widetilde{z_d}}
\newcommand{\tdl}{\Delta\widetilde{ l}}
\newcommand{\tz}{\widetilde{z}}
\newcommand{\ty}{\widetilde{y}}

\newcommand{\ttheta}{\widetilde{\theta}}

\newcommand{\dy}{\Delta y}
\newcommand{\dz}{\Delta z}

\newcommand*{\theequationadd}[1]{%
  \the\numexpr\value{equation}+(#1)\relax
}

\lstdefinelanguage
  [GCC]{C++}
  []{C++}
{
  morekeywords={[2]__attribute__},%
  morekeywords={[3]vector_size,r,DeltaH}%
}
\begin{document}

\title{GPU accelerated image reconstruction in a two-strip J-PET tomograph}

\author{%
  Piotr~Bia\l{}as$^1\thanks{Corresponding author: \texttt{piotr.bialas@uj.edu.pl}}$,
  Jakub~Kowal$^1$,
  Adam~Strzelecki$^1$,
  Tomasz~Bednarski$^1$,
  Eryk~Czerwi\'{n}ski$^1$,
  Aleksander~Gajos$^1$,
  Daria~Kami\'{n}ska$^1$,
  \L{}ukasz~Kap\l{}on$^{1,2}$,
  Andrzej~Kochanowski$^3$,
  Grzegorz~Korcyl$^1$,
  Pawe\l{}~Kowalski$^4$,
  Tomasz~Kozik$^1$,
  Wojciech~Krzemie\'{n}$^1$,
  Ewelina~Kubicz$^1$,
  Pawe\l{}~Moskal$^1$,
  Szymon~Nied\'{z}wiecki$^1$,
  Marek~Pa\l{}ka$^1$,
  Lech~Raczy\'{n}ski$^4$,
  Zbigniew~Rudy$^1$,
  Oleksandr~Rundel$^1$,
  Piotr~Salabura$^1$,
  Neha~G.~Sharma$^1$,
  Micha\l{}~Silarski$^1$,
  Artur~S\l{}omski$^1$,
  Jerzy~Smyrski$^1$,
  Anna~Wieczorek$^{1,2}$
  Wojciech~Wi\'{s}licki$^4$,
  Marcin~Zieli\'{n}ski$^1$ \and
  Natalia~Zo\'{n}$^1$\\
  \address{
  $^1$Faculty of Physics, Astronomy and Computer Science Jagiellonian University,
  30-348 Cracow, Poland}\\
  \address{
  $^2$Institute of Metalurgy and Material Science of Polish Academy of Sciences, 
  30-059 Cracow, Poland}\\ 
\address{
  $^3$Faculty of Chemistry, Jagiellonian University, 30-060 Cracow, Poland }\\
\address{
  $^4${\'S}wierk Computing Centre, National Centre for Nuclear Research,
  05-400~Otwock-{\'S}wierk, Poland}\\
}

\maketitle

\begin{abstract}
  We present a fast GPU implementation of the image reconstruction routine, for
  a novel two strip PET detector that relies solely on the time of flight
  measurements.
\end{abstract}


\section{Introduction}

In this paper we present a GPU implementation of list-mode reconstruction
algorithm of a 2D strip PET. This detector consists of two parallel bars
(strips) of scintillator with a photomultiplier attached to each end~\cite{Moskal:2014sra,Moskal:2014dstrlibrary}.

\begin{figure}[h]
  \centering
  \includegraphics[width=\singlecolumnfigurewidth]{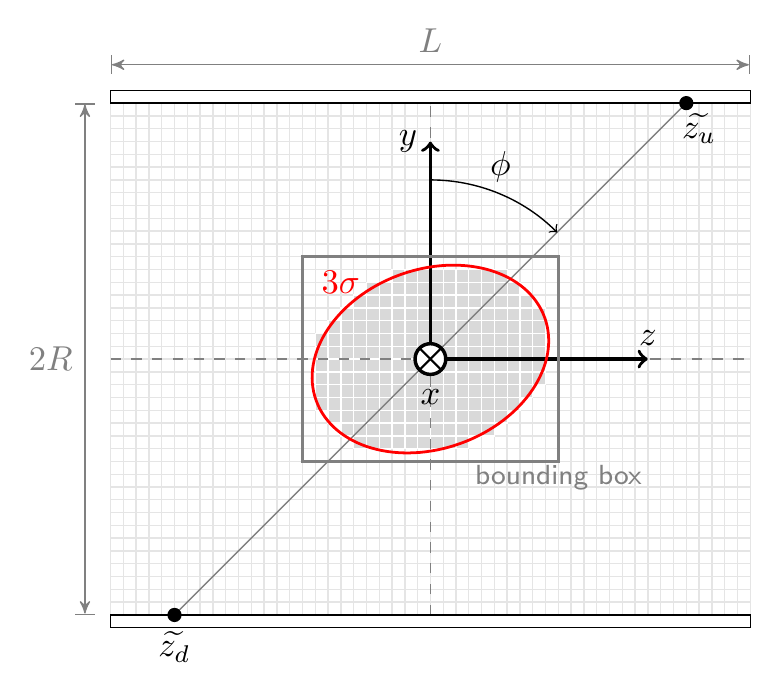}
  \caption{\label{fig:2d}
    2D-strip detector geometry.}
\end{figure}

By measuring the time of the arrivals of photons to each of the
photomultipliers we can reconstruct the position at which $\gamma$ quanta have
interacted with the scintillators as well as the position along the line-of-response (LOR) (see
Figure~\ref{fig:2d}). 
Application of the
state of the art electronics developed at the Jagiellonian University
allowed to achieve the required resolution~\cite{Palka-BAMS,Korcyl-BAMS}.

\begin{figure}
  \centering
  \includegraphics[width=0.9\singlecolumnfigurewidth]{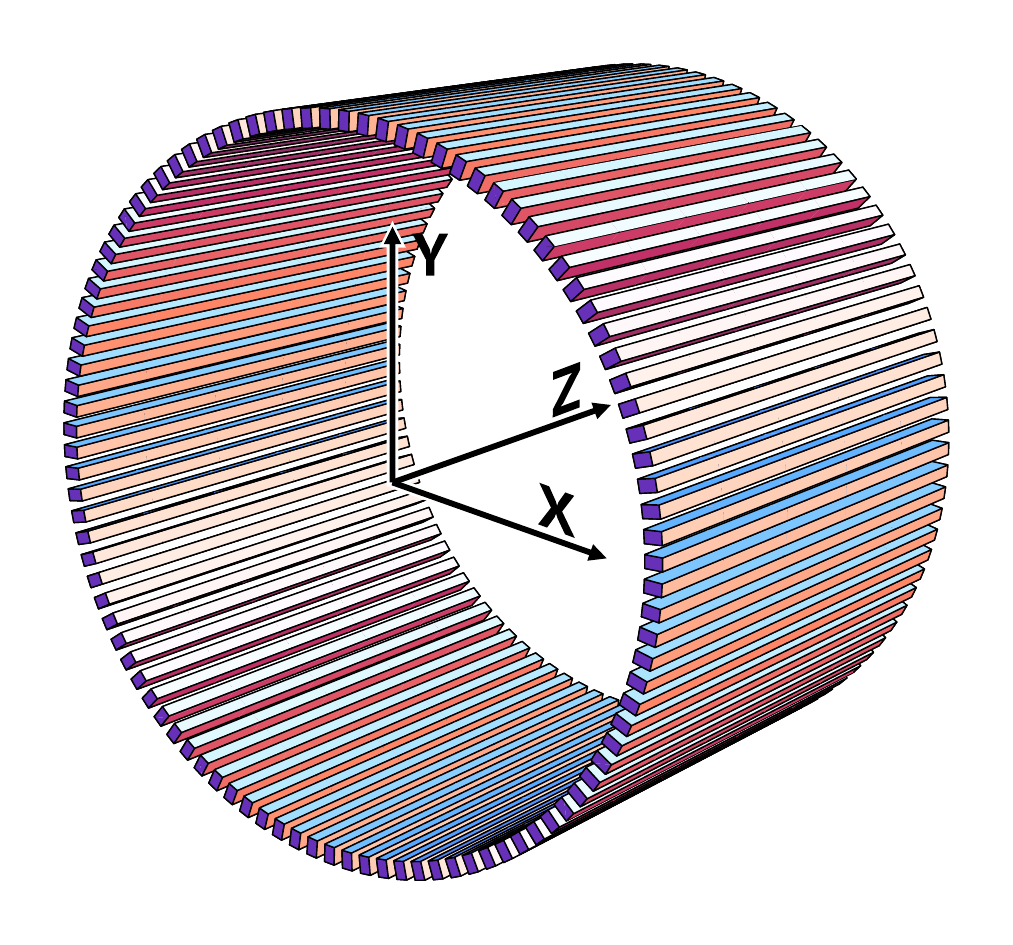}
  \caption{\label{fig:3d} An example of the possible 3D detector geometry of the J-PET detector.}
\end{figure}

A double-strip prototype can be regarded as an elementary part of the full 3D
``J-PET''\footnote{\texttt{http://koza.if.uj.edu.pl/pet/}} detector under
construction at our faculty\cite{Moskal:2014sra,Moskal:2014dstrlibrary,Raczynski:2014poa,organicscint,stripPET}. 
The detector will
consists of cylindrically arranged scintillator strips (as shown schematically in
figure~\ref{fig:3d}) enabling a full 3D reconstruction.
However, the two strip prototype is also of interest as a cheap scanning device.

\section{Setup}

The description of the readout system electronics is beyond the scope of this
paper, we will just assume that for each event we are given three numbers
$(\tzu, \tzd, \tdl)$ (see Fig.~\ref{fig:2d}). By convention we use the
tilde to denote  measured quantities as opposite to "real" or exact values. The $\tzu$ and $\tzd$ denote respectively the
reconstructed position along the upper and lower strip and $\tdl$ is the
difference of the distances along the LOR from the emission point to the upper
and lower strips
\begin{equation}
\begin{split}
\tdl &= \sqrt{(R-\te_y)^2+(\tzu-\te_z)^2} \\
&\phantom{=}-\sqrt{(R+\te_y)^2+(\tzd-\te_z)^2}.
\end{split}
\end{equation}
From those measurements the emission position and angle can be reconstructed directly
\begin{equation}
\begin{aligned}
\tan \ttheta &= \frac{\tzu-\tzd}{2\R}\\
\ty &=-\frac{1}{2}\frac{\tdl}{\sqrt{1+\tan^2\ttheta}}
=\frac{2\R\tdl}{\sqrt{\tzu-\tzd+4\R^2}}\\
\tz&=\frac{1}{2}\left(\tzu+\tzd+2y\tan\ttheta\right)\\
&= \frac{1}{2}\left(
\tzu +\tzd +\frac{(\tzu-\tzd)\tdl}{\sqrt{\tzu-\tzd+4\R^2}}
\right).
\end{aligned}
\end{equation}
It is however subject to measurement errors (see the correlation matrix description at the end of Section~\ref{sec:kernel}).   In 
figure~\ref{fig:direct} we present results of such direct
reconstruction of the phantom depicted in the figure~\ref{fig:true}. It is clear that the resolution of the detector is not sufficient for  direct reconstruction  and statistical reconstruction  methods need to be applied.

The statistical reconstruction is done iteratively using the List-Mode version of the
Maximal Likelihood Expectation Maximization (MLEM) algorithm. Each iteration of
this algorithm defined by the following formula \cite{lmalg}
\begin{equation}\label{eq:iter}
\rho(l)^{(t+1)}=\sum_{j=1}^N
\frac{P(\tev_j|l)\rho(l)^{(t)}}
{\sum\limits_{i=1}^M P(\tev_j|i)s(i)\rho(i)^{(t)}}.
\end{equation}
The $\rho(l)$ is the sought tracer emission density given as the average number
of emissions from pixel $l$ during the examination. The $P(\tev|i)$ is a {\em
reconstruction kernel} that represents the probability that an event
originating in pixel $i$ will be detected as $\tev$. The $s(i)$
is the {\em sensitivity} of the pixel $i$ {\it i.e.} the probability that an
event emitted from pixel $i$ will be detected at all. This sensitivity can be
easily calculated from the geometry:
\begin{equation}\label{eq:s-yz}\begin{split}
s(y,z)
&=\pi^{-1}\biggl(
\arctan\min\biggl(\frac{\frac{1}{2}L-z}{R-y},\frac{\frac{1}{2}L+z}{R+y}\biggr)\\
&\phantom{=-\pi^{-1}}-
\arctan\max\biggl(-\frac{\frac{1}{2}L+z}{R-y},\frac{-\frac{1}{2}L+z}{R+y}\biggr)\biggr).
\end{split}
\end{equation}
In derivation have assumed the detection probability along the strip is
constant and that it does not depend on the angle of incidence. This conditions
should be approximately fulfilled for incidence angles not exceeding $30^\circ$.

The formula \eqref{eq:iter}  can be rewritten as
\begin{equation}\label{eq:iter-p}
\rho'(l)^{(t+1)}=\sum_{j=1}^N
\frac{P(\tev_j|l)\rho'(l)^{(t)}}
{\sum\limits_{i=1}^M P(\tev_j|i)\rho'(i)^{(t)}}.
\end{equation}
with
\begin{equation}
\rho'(i)\equiv s(i)\rho(i).
\end{equation}
In the following we will give the results of the reconstruction of $\rho'(i)$.

The sum over $j$ in \eqref{eq:iter-p} runs over all collected events $\{\te_j\}$.
Considering that up to hundred millions of events can be collected during a
single scan this is a very time consuming calculation so the efficient
calculation of the kernel $P$ is essential.

\section{Kernel and correlation matrix}
\label{sec:kernel}

In \cite{skernel,lmpet} we have found analytical approximation of $P(\tev|i)$ given by
\begin{equation}\label{eq:kernel}
\begin{split}
  P(\tev|i)
  &\approx\frac{\det^{\frac{1}{2}} C }
               {2\pi\sqrt{\vec{a}C^{-1}\vec{a}+2\vec{o}C^{-1}\vec{b}}}
\\
&\phantom{\approx}
\exp\left(-\frac{1}{2}\left(\vec{b} C^{-1}\vec{b}
-\frac{\left(\vec{b}C^{-1}\vec{a}\right)^2}
      {\vec{a}C^{-1}\vec{a}+2\vec{o}C^{-1}\vec{b}}\right)\right)
\end{split}
\end{equation}
The $\vec{o}$, $\vec{a}$, $\vec{b}$ are defined as follows
\begin{align}
\vec{o}&=\begin{pmatrix}
-(\dy +\ty -R)\tan\ttheta\cos^{-2}\ttheta\\
-(\dy +\ty +R)\tan\ttheta\cos^{-2}\ttheta\\
-(\dy+\ty)\cos^{-1}\ttheta(1+2\tan^2\ttheta)
\end{pmatrix},\\
\vec{a}&=\begin{pmatrix}
-(\dy +\ty -R)\cos^{-2}\ttheta\\
-(\dy +\ty +R)\cos^{-2}\ttheta\\
-(\dy+\ty)\cos^{-1}\ttheta\tan\ttheta
\end{pmatrix},\\
\vec{b}&=
\begin{pmatrix}
\dz-\dy \tan\ttheta\\
\dz-\dy \tan\ttheta\\
-2\dy \cos^{-1}\ttheta
\end{pmatrix}
\end{align}
and
\begin{equation}
\dy=y-\ty\quad\text{and}\quad \dz=z-\tz.
\end{equation}
The $\ty$ and $\tz$ are the coordinates of the reconstructed emission point and
$\ttheta$ is the reconstructed emission angle of the event $\tev$. The $y$ and
$z$ are the coordinates of the center of pixel $i$. $C$ is the correlation
matrix which in general can be of the form:
\begin{equation}
C^{-1}=
\begin{pmatrix}
\frac{1}{\sigma_z^2} & 0  &  \gamma \\
0 &     \frac{1}{\sigma_z^2} & -\gamma\\
\gamma & -\gamma   & \frac{1}{\sigma_{\dl}^2}
\end{pmatrix}.
\end{equation}
This matrix depends on the $\tzu$ and $\tzd$. For $\sigma_z$.  
Experimentally we have found this dependence to be quite weak 
on the order of 10\%   from the center (lowest) to the edge (highest). 
 We have found out that coefficient $\gamma$ 
 can be neglected  as long as we do not take into account events with $z_{u(d)}$ near the edge of the scintillators. This may change when we consider the full detector with longer (500mm) scintillator strips, but  in this contribution we assume correlation matrix to be diagonal. Currently we achieve $\sigma_{z}\approx10mm$ 
and $\sigma_{\dl}\approx40mm$.  
Please note that the last number corresponds to $20mm$ error
for the position along the emission line as the distance from the line midpoint is equal to $\frac{1}{2}\dl$.

Formula \eqref{eq:kernel} is, at least for the range of parameters we have
studied, strongly dominated by the gaussian term $\vec{b} C^{-1}\vec{b}$. This
term defines an $3\sigma$ ellipse (see figure~\ref{fig:2d}). For practical
purposes we can assume that the kernel is zero outside this ellipse. As it is
easier to work with rectangular shapes we also define a bounding box consisting
of an rectangle that is circumscribed on the ellipse (see
Appendix~\ref{app:bounding-box}).


\section{Implementation}

The iteration step described by formula \eqref{eq:iter} can be implemented
as described by the pseudocode in Listing~\ref{lst:mldm}.

\lstset{
  basicstyle=\ttfamily\small,
  keywordstyle=[2]\textit,%
  morekeywords={[2]p,s,ellipse},%
}
\begin{lstlisting}[frame=single,
  caption=Implementation of the reconstruction iteration routine.,
  label={lst:mldm},
  frame=none,
  float]
for (auto p_l : pixels) {
  rho_new[p_l] = 0.0;
}
for (auto e_j : events) {
  auto denominator = 0.0;
  for (auto i : ellipse(e_j)) {
    kernel[i] = p(e_j, i);
    denominator += kernel[i] * rho[i];
  }
  for (auto i : ellipse(e_j)) {
    rho_new[i] += rho[l] * kernel[i] / denominator;
  }
}
\end{lstlisting}

Loops \lstinline!for(auto i : ellipse(e_j))! on lines $6$ and $10$ iterate over
all pixels in the $3\sigma$ ellipse of the event $\tev$. To calculate pixels
contributing to this ellipse we first need to determine its bounding box in
pixel space. Once bounding box is calculated we loop only trough pixels inside
this bounding box. Each pixel is then tested if its center point resides inside
or outside of the ellipse. Only then the whole kernel is calculated. The
results are cached and used subsequently in the second loop.

\subsection{CPU}

The CPU implementation follows essentially the algorithm from
listing~\ref{lst:mldm}. We use OpenMP to parallelize the outer loop (line 5)
over the events. Each thread writes to its own copy of \lstinline!rho_new!
array which are added at the end of the iteration. Currently we do not take
direct advantage of the AVX/SSE instruction set aside of automatic
vectorization provided by \emph{Intel C++ Compiler}.

\subsection{GPU implementation}

Next step was a naive GPU implementation based on our reference CPU
implementation where each thread processes all pixels of single event, so few
thousands of events are processed simultaneously by hardware threads.

Such approach has however serious drawback on GPU hardware, which is
essentially a vector computer.  On the NVIDIA CUDA architecture that we use, the threads are collected in batches of 32
threads called {\em warps}. All threads in a warp must execute same instruction
in parallel (SIMD). In the naive implementation each thread is processing a
different events with different number of pixels. That amounts to a double loop
with loops bounds different across the threads of a warp. This leads to severe
{\em thread divergence} and as we have discovered carries a much higher penalty
then naively expected. One would expect that the execution time of a warp,
would be approximately the time needed to execute the longest loops, but as it
turned out it is much higher. Additionally we cannot cache visited pixels and
their kernel results since it is not enough registers or shared memory to store
such information given each thread processes separate event.

\begin{figure}[t]
  \centering
  \includegraphics[width=\singlecolumnfigurewidth]{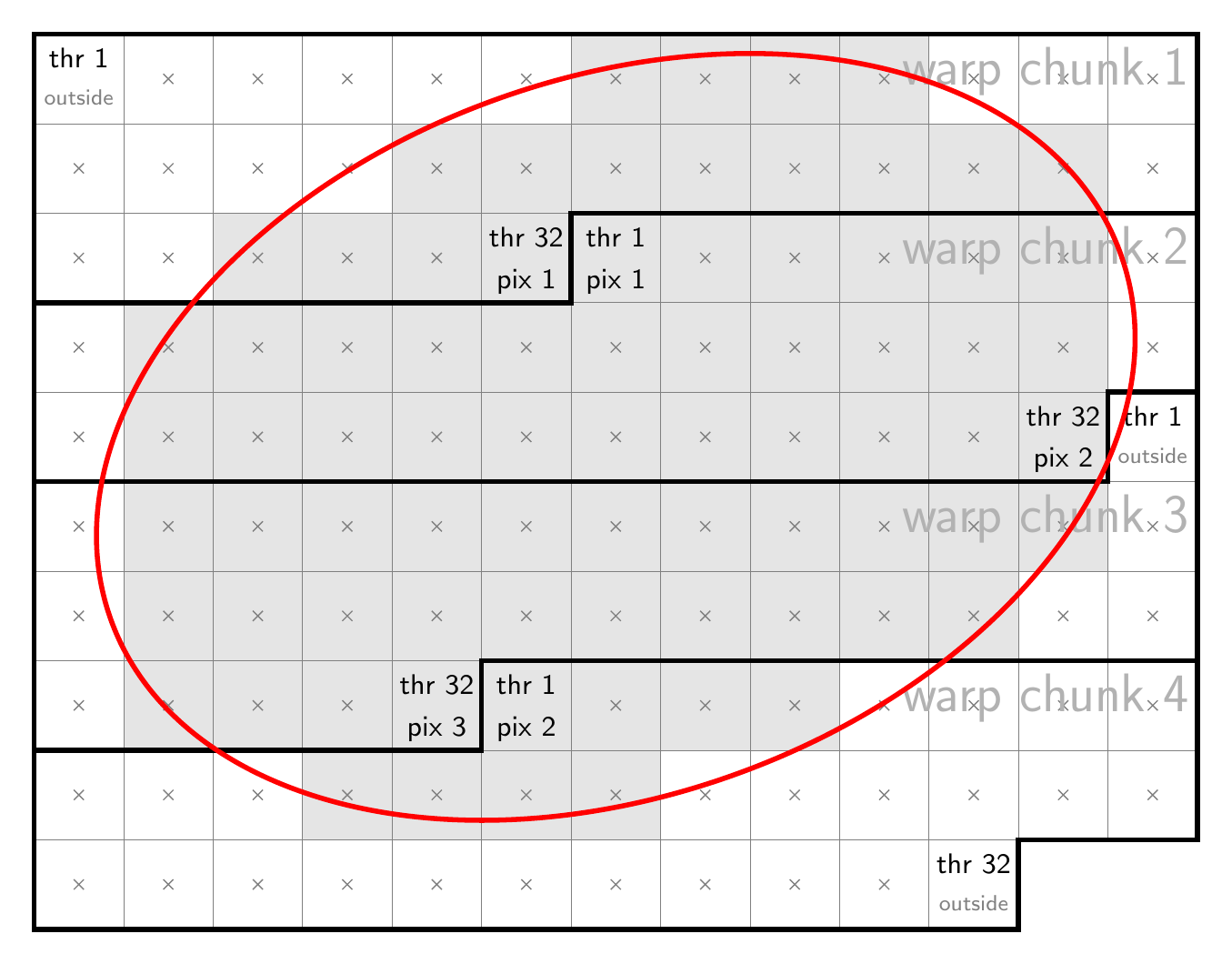}
  \caption{\label{fig:warp-granularity}
 Warp granularity \small (whole event processed by single warp)}
\end{figure}

We can circumvent that using different pixel calculation scheduling where whole
warp of 32 threads calculates a single event. This is called by us \emph{warp
granularity} (see Figure~\ref{fig:warp-granularity}). As each thread in a warp process a single pixel from the same event there is no
divergence. Different events are processed by different warps which can run
independently. This algorithm also lets us better leverage
available shared memory and registers. However, processor cycles are still wasted by the threads that
fail the bounding ellipse test. 

First it has to be noted that single event is calculated in two passes. First
we need to calculate denominator of \eqref{eq:iter}. This pass needs bounding box to be calculated first,
then each pixel in this pass is tested with \emph{3-sigma} ellipse equation.

During first pass \emph{warp granularity} gives us opportunity to cache visited
pixels and kernel \eqref{eq:kernel}  in shared memory and registers, so the
second pass can loop only through visited already pixels without a need to test
them for ellipse inclusion. Also we can cache kernel results in registers,
since each thread in warp is likely to visit just few pixels of single event.

Calculation of the denominator requires adding the contributions from the 32 threads of the warp.
We have done this using the  new shuffle instructions introduced in Kepler architecture. 
This gave a notable performance boost over standard  reduction algorithm using shared memory\cite{Harris}.

Final optimization is to access $\rho$ (previous iteration image buffer) as
texture. This produces noticeable performance boost by using hardware GPU
texture unit cache and special 2D access optimized memory layout.  
However it 
can be observed that memory
access still takes around $35\%$ of overall iteration time after optimizations.

\section{Benchmarks and results}

We have benchmarked our GPU implementation on \emph{NVIDIA GeForce} GTX 770
commodity card with 4GB memory and compute capability 3.0 using \emph{CUDA} SDK
$6.5$, CPU implementation on \emph{Intel Xeon} CPU E5-1650 v2 @ $3.50$GHz with
6 cores using \emph{ICC} $15.0.0$ (\emph{Intel Composer XE} $2015$). The benchmark results are presented in Table~\ref{tbl:benchmark}, while 
the results of reconstruction of the phantom after different number of
iterations are presented in the figures~\ref{fig:i1} to \ref{fig:i4}.

\begin{table}[ht!]
\centering
  \newcolumntype{R}{>{\centering\arraybackslash}p{3em}}
  \newcolumntype{S}{>{\centering\arraybackslash}p{4em}}
  \newcolumntype{N}{>{\raggedleft\arraybackslash}p{4em}}
  \begin{filecontents*}{time-1iteration-per-events.txt}
# it        cpu   gpu-warp gpu-thread speedup
  10   11827.69     465.95     689.25      25
  20   23653.35     931.11    1376.29      25
  30   35455.40    1397.72    2066.04      25
  40   47200.46    1863.60    2753.05      25
  50   58991.09    2329.90    3440.05      25
  60   70815.31    2795.80    4128.14      25
  70   82726.57    3261.03    4817.00      25
  80   94411.87    3727.25    5502.56      25
  90  106148.54    4194.00    6189.43      25
 100  118040.53    4659.73    6876.62      25
\end{filecontents*}
\pgfplotstabletypeset[
    begin table/.add={}{[.7\textwidth]},
    columns={0,1,3,2,4},
    columns/0/.style={
      column name=Number of~Events,
      column type={|N|},
      postproc cell content/.style={@cell content/.add={}{$\times10^6$}}},
    time/.style={
      column type={r|},
      preproc/expr={##1/1000},
      fixed, fixed zerofill=true, precision=2,
      postproc cell content/.style={@cell content/.add={}{ s}}}, 
    columns/1/.style={time, column name=CPU \scriptsize OpenMP, column type=S|},
    columns/3/.style={time, column name=GPU \scriptsize Thread, column type=R|},
    columns/2/.style={time, column name=GPU \scriptsize Warp, column type=R|},
    columns/4/.style={column name=Speedup \scriptsize CPU/Warp, column type=S|,
      postproc cell content/.style={@cell content/.add={}{$\times$}}},
    every head row/.style={before row=\hline},
    every first row/.style={before row=\hline},
    every last row/.style={after row=\hline},
  ]{time-1iteration-per-events.txt}
\caption{\label{tbl:benchmark}Single iteration reconstruction time per number of events.}
\end{table}

\begin{figure}
\centering
\subfloat[\label{fig:true}Ideal reconstruction]{
  \includegraphics[width=.65\singlecolumnfigurewidth]{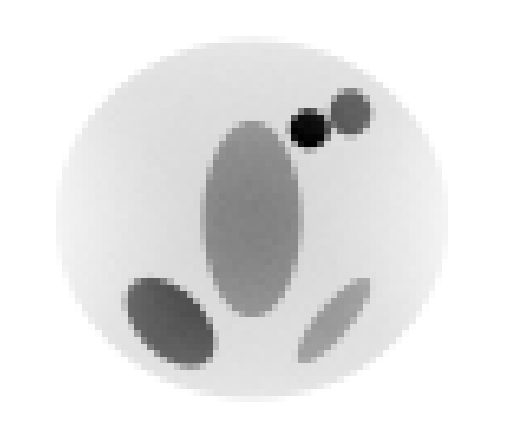}}
\subfloat[\label{fig:direct}Direct reconstruction]{
  \includegraphics[width=.65\singlecolumnfigurewidth]{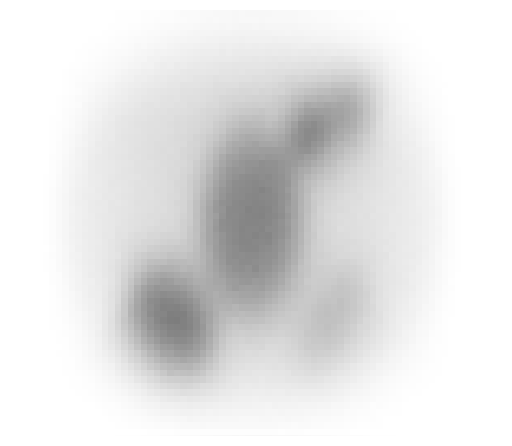}}
\\
\subfloat[\label{fig:i1}after 1 iteration]{
  \includegraphics[width=.65\singlecolumnfigurewidth]{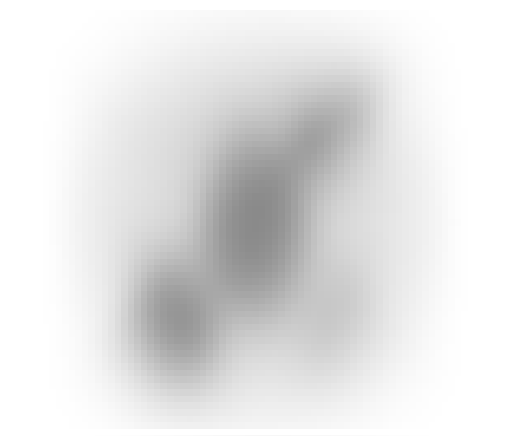}}
\subfloat[\label{fig:i2}after 5 iterations]{
  \includegraphics[width=.65\singlecolumnfigurewidth]{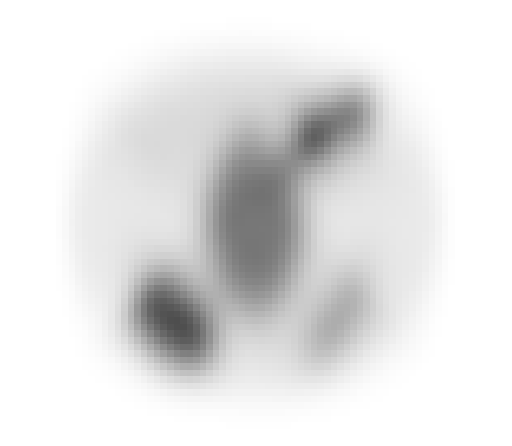}}
\\
\subfloat[\label{fig:i3}after 10 iterations]{
  \includegraphics[width=.65\singlecolumnfigurewidth]{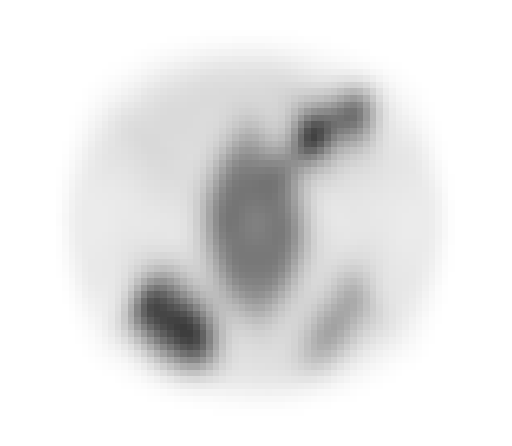}}
\subfloat[\label{fig:i4}after 25 iterations]{
  \includegraphics[width=.65\singlecolumnfigurewidth]{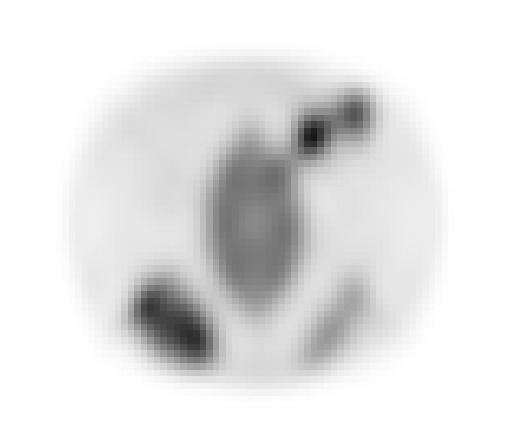}}
\caption{Phantom used in reconstruction. 
         ($R=130$mm, $L=300$mm and $4\times4$mm pixel size)}
\end{figure}


\section{Summary and outlook}

We have implemented and tested our reconstruction kernel on simulated  data  using realistic parameters obtained from experimental measurements.  As seen from  the figures~\ref{fig:i1} to \ref{fig:i4} the results are very encouraging, considering the simplicity and the resolution of our setup. Implementing the reconstruction algorithm on the commodity  GPU  provided a 25-fold speedup that allows real-time processing. 
One should note however that this speed is partly due to not taking advantage of the CPU vector AVX instruction set. The reason for this is that as we have already pointed out in \cite{bks} the CUDA and OpenCL programming model is inherently vectorized while CPU is still viewed as superscalar processor with vector instructions mixed in. This is  only now slowly changing with introduction of new compiler pragmas to deal explicitly with vectorization in a similar way as OpenMP deals with parallelization.  

In derivation of the \eqref{eq:kernel} we have assumed a very simple detector geometry with scintillators approximated by  thin lines.   In reality they have a rectangular cross section of 5x20$mm^2$. To some extent this was taken into account by using the errors estimated from real scintillators. The model however must be validated on real data (which is already collected) and this is a subject of an ongoing  investigation.

\section*{Acknowledgments}

We acknowledge technical and administrative support by 
T. Gucwa-Ry\'{s}, 
A. Heczko,
M. Kajetanowicz, 
G. Konopka-Cupia\l{}, 
W. Migda\l{}, 
and the financial support
by the Polish National Center for Development and Research 
through grant No. INNOTECH-K1/IN1/64/159174/NCBR/12,
the Foundation for Polish Science through MPD programme,
the EU, MSHE Grant No. POIG.02.03.00-161 00-013/09, 
and Doctus - the Lesser Poland PhD Scholarship Fund.


\appendix

\section{Phantom}

Phantom definition is given in table~\ref{tab:phantom}. Each row corresponds to
an ellipse with center $(x,y)$ the half-axes $a$ and $b$ rotated by angle
$\phi$ counter-clockwise. The $\rho$ denotes the relative density of the
tracer. When two ellipses overlap the $\rho$ is taken from the topmost (in
table) one.

\begin{table}
\begin{center}
\begin{tabular}{|r|r|r|r|r|r|}
\hline
  x  &    y  &    a  &    b & $\phi$ & $\rho$ \\\hline
  0  &    0  &   30  &   60 &     0  &    0.3 \\
 50  &  -62  &   10  &   33 &   -40  &    0.3 \\
-50  &  -63  &   20  &   33 &    45  &    0.5 \\
 60  &   65  &   13  &   14 &     0  &    0.5 \\
 35  &   55  &   12  &   12 &     0  &    0.7 \\
  0  &    0  &  120  &  110 &     0  &    0.1 \\
\hline
\end{tabular}
\end{center}
\caption{\label{tab:phantom} Phantom description.}
\end{table}

\section{Bounding box}
\label{app:bounding-box}

Given an ellipse defined by the equation
\begin{equation}
Ay^2+C y z + B z^2 = R^2
\end{equation}
its bounding box is a rectangle with lower left corner at
\begin{equation}
y=-\frac{R}{\sqrt{A-\frac{C^2}{B}}}, \qquad z=-\frac{R}{\sqrt{B-\frac{C^2}{A}}}
\end{equation}
and symmetric upper right corner. This combined with
\begin{equation}\begin{split}
\vec{b}C^{-1}\vec{b} & =
2 \frac{(\dz -\dy \tan\ttheta)^2}
       {\sigma_z^2}+4\frac{\dy^2}{\sigma_{\dl}^2cos^2\ttheta} \\
& = \dz^2\frac{2}{\sigma_z^2} - 2\dz\dy\frac{2\tan\ttheta}{\sigma_z^2}\\
&\phantom{=\dz^2} + \dy^2\left(\frac{2\tan^2\ttheta}{\sigma_z^2}
                  + \frac{4}{\sigma_{\dl}^2\cos^2\ttheta}\right)
\end{split}
\end{equation}
allows us to calculate the bounding box of the $3\sigma$ ellipse for each event.

\end{document}